\documentclass[useAMS,usenatbib]{mn2e}
\usepackage{epsfig}
\usepackage{amssymb}
\usepackage{amsmath}
\usepackage{natbib}
\usepackage{threeparttable} 
\usepackage{graphicx}
\usepackage{setspace}
\usepackage{longtable}
\usepackage{multicol}
\usepackage{subfigure}

\newcommand\apj{ApJ}
\newcommand\apjs{ApJS}
\newcommand\aap{A\&A}
\newcommand\aaps{A\&AS}
\newcommand\aj{AJ}
\newcommand\mnras{MNRAS}
\newcommand\pasp{PASP}
\newcommand\pasj{PASJ}
\newcommand\gca{Geochim.~Cosmochim.~Acta}
\newcommand\ssr{Space Sci. Rev.}%
\newcommand\memsai{MnSAI}

\def\hide#1{}

\newcommand{\ax}{AX J1549.8--5416}

\newcommand{\chan}{\textit{Chandra}}
\newcommand{\swift}{\textit{Swift}}

\newcommand{\xmm}{\textit{XMM-Newton}}

\newcommand{\Msun}{\mathrm{M}_{\odot}}

\newcommand{\cnts}{\mathrm{counts~s}^{-1}}

\def\mathnew{\mathsurround=0pt}
\def\simov#1#2{\lower .5pt\vbox{\baselineskip0pt \lineskip-.5pt
        \ialign{$\mathnew#1\hfil##\hfil$\crcr#2\crcr\sim\crcr}}}

\def\simless{\mathrel{\mathpalette\simov <}}

\title[A new dwarf nova AX J1549.8--5416]{Multiwavelength monitoring of a very active dwarf nova AX J1549.8$-$5416 
with an unusually high duty cycle}

\author[G. Zhang et al.]{Guobao Zhang$^{1}$\thanks{E-mail: guobao.zhang@nyu.edu}, 
Joseph D. Gelfand$^{1, 2}$, David M. Russell$^{1}$, Fraser Lewis$^{3, 4}$
\newauthor
Nicola Masetti$^{5, 6}$, Federico Bernardini$^{1}$, Ileana Andruchow$^{7, 8}$ and L. Zibecchi$^{7, 8}$ \\
$^{1}$New York University Abu Dhabi, P.O. Box 129188, Abu Dhabi, UAE \\
$^{2}$Center for Cosmology and Particle Physics, New York University, Meyer Hall of Physics, 
4 Washington Place, New York, NY 10003 \\
$^{3}$Faulkes Telescope Project, School of Physics and Astronomy, Cardiff University, 5 The Parade, Cardiff, CF24 3AA, Wales, UK \\
$^{4}$Astrophysics Research Institute, Liverpool John Moores University, IC2, Liverpool Science Park, 146 Brownlow Hill, Liverpool L3 5RF, UK \\ 
$^{5}$INAF $-$ Istituto di Astrofisica Spaziale e Fisica Cosmica di Bologna, via Gobetti 101, 40129 Bologna, Italy \\
$^{6}$Departamento de Ciencias F\'isicas, Universidad Andr\'es Bello, Fern\'andez Concha 700, Las Condes, Santiago, Chile \\
$^{7}$Facultad de Ciencias Astron\'omicas y Geof\'isicas, Universidad Nacional de La Plata, Paseo del Bosque, B1900FWA La Plata, Argentina \\
$^{8}$Instituto de Astrof\'isica La Plata (IALP), CONICET-UNLP, Argentina
}

\begin{document}

\maketitle

\label{firstpage}


\begin{abstract}
We present the results of our analysis of new optical, ultraviolet (UV) 
and X-ray observations of a highly variable source $-$ \ax. Both the detection 
of several fast rise, exponential decay outbursts in the optical light curve
and the lack of  He II emission lines in the optical spectra  suggest \ax\ 
is a cataclysmic variable of the dwarf nova (DN) type. The multiwavelength 
analysis of three mini-outbursts and one normal outburst represent one of 
the most complete multiwavelength studies of a DN and help to refine the 
relationship between the X-ray, UV and optical emission  in this system. 
We find that the UV emission is delayed  with respect to the optical by 
$1.0-5.4$ days during the rising phase of the outburst. The X-ray emission 
is suppressed during the peak of the optical outburst  and recovers during 
the end of the outburst.  From  our analysis of archival \swift, \chan\ and \xmm\ 
observations  of \ax,  we estimate this DN has a high duty cycle ($\sim50\%$), 
suggesting a quiescent  X-ray luminosity larger than $10^{32}$ erg/s. 
We also find  the X-ray and UV flux are roughly anti-correlated.
Furthermore, we find that, at low X-ray fluxes, the X-ray spectrum is well 
described by a single temperature thermal plasma model, while at high 
X-ray fluxes,  an isobaric cooling flow model also works. We find that the 
maximum temperature of the plasma in quiescence is significantly higher than 
that in outburst.

\end{abstract}


\section{introduction}
\label{introduction}

Cataclysmic variables (CVs) are interacting binaries consisting of a low-mass 
secondary star very close to a more massive white dwarf (WD) primary star. 
The WD accretes matter from  the companion star via Roche lobe overflow.  
CVs are classified into several types based on their observational characteristics, 
depending on the nature of the primary star and the accretion process 
\citep[see reviews by ][]{Kuulkers06, Singh13}. In particular,
CVs are divided into non-magnetic CVs (non-MCVs) and magnetic CVs (MCVs) 
based on the strength of the WD surface magnetic field. In non-MCVs, X-rays are 
thought to originate from the boundary layer (BL) between the slowly rotating 
accreting WD and the fast rotating (Keplerian) inner edge of the accretion disc, 
where the material dissipates its remaining rotational kinetic energy before
accreting onto the surface of the WD. In MCVs, X-rays originate from an accretion 
column at or near the magnetic poles. The columns may be fed via magnetospheric 
accretion from the inner boundary of a truncated disc, as in the subclass of 
intermediate polars (IPs), or from a funneled accretion stream in polars. The X-ray 
spectra of magnetic and non-magnetic CVs are typically  described by multi-temperature 
thermal plasma emission \citep[e.g., ][]{Kuulkers06, Singh13}.

Dwarf novae (DNe) are non-MCVs with an accretion disc that flows onto the WD primary, 
and a secondary star.  The orbital periods of DNe are typically between 70 minutes
and 10 hours \citep{Kuulkers06, Patterson11, Balman12}. In DNe, the accretion disc 
has two stable states which correspond to optical quiescence and outbursts. 
During optical quiescence, the X-ray emission arises from the hot, optically thin, 
thermal plasma produced at the  BL.  The X-ray spectrum is hard, 
has a temperature greater than 10 keV, and typically features associated with the 
Fe K$_{\alpha}$ line are observed. Furthermore, during this state the optically 
thick accretion disc is truncated, and the accretion rate is less than $10^{-9}-10^{-9.5}$ $\Msun$/year.

When in outburst, the optical flux increases by 2$-$9 magnitudes for several days 
to  weeks. This increase is associated with an increase in the luminosity of the 
accretion disc which dominates the emission at optical and UV wavelengths 
\citep[e.g., ][]{Lasota01, Kuulkers06}. During this period, the X-ray spectra is 
soft, and the  plasma temperature decreases such that the thermal emission 
peaks in the extreme UV. Due to the increase in accretion rate on the WD,  the 
plasma in the BL becomes optically thick and therefore more efficient at cooling
\citep{Pandel03, Saitou2012, Baskill05, Collins10, McGowan04}. The mechanism leading 
to repeated transitions between the optical outburst and quiescent states is 
thought to be thermal viscous disc instability \citep[e.g., ][]{Meyer81, Lasota01}.

An analysis of archival \swift\  data identified a highly variable X-ray source --
\ax\ -- which was first detected in the ASCA Galactic plane Survey \citep{Sugizaki01}.
Its optical counterpart was proposed to be NSV 20407, a variable star 
with a B-band magnitude ranging from $> 18$ to 16.7 \citep{Samus09}, suggesting this 
source is a CV.  \cite{Lin14} analysed four \xmm\ observations of 
\ax\ and found that the source's X-ray flux varied considerably.
They also detected a Fe line in the X-ray spectra of two 
observations and suggested this source be a  MCV, which often have
Fe emission in their X-ray spectra \citep{Ezuka99}.

In order to further investigate the nature of \ax, we observed the source at optical 
wavelengths with the Las Cumbres Observatory Global Telescope Network 
(LCOGT)  and 'Jorge Sahade' at CASLEO (Argentina), complemented with UV 
and X-ray observations by \swift.  Since  \ax\ is located $\sim 9.2$ arcmin 
away from 1E 1547.0$-$5408, it is within the field of view of a large number of 
archival \swift\ observations. In this paper,  we also analysed all available 
archival \xmm, \chan, and \swift\ data of the source. We describe the observations 
and  data reduction in Section \ref{data}, and we present our results in Section 
\ref{result}. Finally, in Section \ref{discussion} we discuss our findings and 
summarize our conclusions in Section \ref{conclusion}.

\section{Observations and data reduction}
\label{data}

\subsection{LCOGT Optical observations} 
\label{optical data}

We observed the field of AX~J1549.8--5416 with the LCOGT suite of robotic 1-m telescopes, 
from  2015-04-30 to 2015-06-18  and from 2016-02-01 to 2016-04-30. 
Some data in 2016 were taken with the 2-m Faulkes Telescope South (FTS).
These data were acquired as part of our Faulkes Telescopes monitoring project of 
X-ray transients \citep{Lewis08}. Optical images were acquired in four filters: Bessel 
$B$-band, $V$-band,  $R$-band  and SDSS $i^{\prime}$-band. The observation 
logs are provided in Table \ref{tab:optical}.  All LCOGT telescopes are 
equipped with a camera with pixel scale 0.467 arcsec pixel$^{-1}$, 
while the camera of FTS has 0.301 arcsec pixel$^{-1}$. Bias subtraction and flat-fielding
were performed via the automatic 
pipelines\footnote{https://lcogt.net/observatory/data/BANZAIpipeline}.

\begin{figure}
    \centering
        \includegraphics[width=3.1in,angle=0]{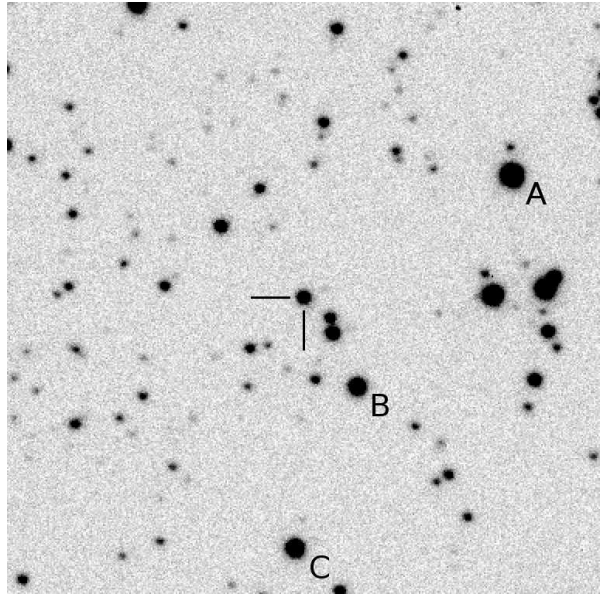}
    \caption{Optical ($R$-band, taken on 2015-06-10 at Cerro Tololo) finding 
    chart. The field of view is 3.0 $\times$ 3.0 arcmin; north is up and east 
    is to the left. AX J1549.8--5416 is marked at the centre, and the stars 
    A, B and C are the three used for flux calibration.}
    \label{fig finding chart}
\end{figure}

The optical counterpart of \ax\ is clearly visible, 
at a location consistent with the position of the X-ray  
source, the UV source and the known variable star NSV 20407. Photometry was 
performed on the target and three field stars in the APASS data release 9 
catalogue \citep[][see Fig. \ref{fig finding chart}]{Henden15} using 
\small PHOT \normalsize in \small IRAF\normalsize, adopting a 2.8 arcsec 
aperture. We calibrated the field using the catalogue $B$, $V$ and $i^{\prime}$-band 
magnitudes of the three field stars, which have typical errors of 0.01--0.1 mag. 
$R$-band flux calibration was achieved using the transformation from $V$, $r^{\prime}$ 
and $i^{\prime}$-bands described by \cite{Jordi06}.

\begin{table*}
\begin{center}
\caption{New  \swift/XRT observations. The error bars are given in 1-$\sigma$ level.}
\vspace{-2mm}
\begin{tabular}{ccccccc}
\hline
ObsID       & Date UT             &       MJD      & exposure (s) & X-ray rate ($\times10^{-3}$cnt/s) &  UV magnitude &  UVOT filter \\
\hline
00042706002 & 2016-02-11T17:40:32 & 57429.73648148 & 685  & 6.09 $\pm$ 3.04  & $>$ 20.3         & uvw2 \\
00042706003 & 2016-02-15T22:14:32 & 57433.92675926 & 505  & 4.19 $\pm$ 2.96  & 19.37 $\pm$ 0.2  & uvw2 \\
00042706004 & 2016-02-19T15:25:11 & 57437.64248843 & 585  & 31.61 $\pm$ 7.45 & 19.31 $\pm$ 0.18 & uvw2 \\
00042706005 & 2016-02-24T22:58:47 & 57442.95748843 & 850  & 16.94 $\pm$ 4.52 & 19.22 $\pm$ 0.18 & uvm2 \\
00042706006 & 2016-02-27T11:48:53 & 57445.49228009 & 775  & 13.17 $\pm$ 4.16 & 20.04 $\pm$ 0.29 & uvw2 \\
00042706007 & 2016-03-04T04:42:00 & 57451.19583333 & 1135 & 7.19 $\pm$ 2.54  & 19.11 $\pm$ 0.12 & uvw1 \\
00042706008 & 2016-03-14T20:13:22 & 57461.84322042 & 995  & 3.08 $\pm$ 1.78  & 19.21 $\pm$ 0.13 & uvw2 \\
00042706009 & 2016-03-18T04:11:02 & 57465.17493410 & 810  & 1.25 $\pm$ 1.25  & 18.89 $\pm$ 0.13 & uvw2 \\
00042706010 & 2016-03-25T08:07:36 & 57472.33861111 & 1020 & 13.11 $\pm$ 3.63 & 18.50 $\pm$ 0.15 & u \\
00042706011 & 2016-03-28T01:36:03 & 57475.06670139 & 895  & 11.31 $\pm$ 3.57 & 19.04 $\pm$ 0.14 & uvw1 \\
00042706012 & 2016-03-31T21:52:52 & 57478.91171296 & 965  & 22.07 $\pm$ 4.81 & $>$ 20.28        & uvm2 \\
00042706013 & 2016-04-03T01:12:19 & 57481.05021991 & 1000 & 19.41 $\pm$ 4.45 & $>$ 20.37        & uvw2 \\
00042706014 & 2016-04-06T12:13:08 & 57484.50912037 & 890  & 6.92 $\pm$ 2.82  & $>$ 20.41        & u) \\
00042706015 & 2016-04-09T10:27:51 & 57487.43600694 & 485  & 2.07 $\pm$ 1.99  & 16.33 $\pm$ 0.04 & uvw1 \\
00042706016 & 2016-04-10T19:56:28 & 57488.83167344 & 472  & 6.35 $\pm$ 3.67  & 15.82 $\pm$ 0.03 & u \\
00042706016 &       ---           &     ---        & ---  &     ---          & 16.88 $\pm$ 0.09 & uvw2 \\
00042706017 & 2016-04-16T22:34:05 & 57494.94112905 & 830  &     ---          & 17.94 $\pm$ 0.08 & uvm2 \\
00042706018 & 2016-04-19T22:27:44 & 57497.93672437 & 945  &     ---          & 18.85 $\pm$ 0.20 & uvw2 \\
00042706019 & 2016-04-22T08:58:27 & 57500.37471492 & 904  & 9.60 $\pm$ 3.32  & 18.17 $\pm$ 0.06 & u \\
00042706020 & 2016-04-25T23:17:06 & 57503.97099751 & 670  & 10.14 $\pm$ 3.95 & 19.86 $\pm$ 0.29 & uvw1 \\
\hline
\end{tabular}
\normalsize
\label{tab:xrt}
\end{center}
\end{table*}

For the first four observations, all four filters were used to constrain the colour 
and variability amplitude of the source. Variations of amplitude $\sim 2$ 
mag were seen in all bands, and for subsequent observations just $R$-band was used 
(see Table \ref{tab:optical} for details). As a check of our 
relative errors, photometry was performed on a fourth field star in $R$-band. 
Its magnitude differed from its mean value of $R = 16.11$ mag by $< 2 \sigma$ 
in 25 of 26 images in 2015. The mean error on the magnitude of this field star measured 
from each image was 0.03 mag, and for AX~J1549.8--5416 the mean magnitude 
errors were 0.10, 0.05, 0.04 and 0.04 mag in $B$, $V$, $R$ and $i^{\prime}$-bands, 
respectively. The image quality was particularly bad in the observation 
taken on 2015-06-18, which resulted in a larger magnitude error. On 
2015-06-10, 20 $R$-band images were taken to search for short-term 
variability. The results are shown in Section \ref{optical}. 
In 2016 we carried out a high cadence monitoring campaign in R-band to
closely follow the morphology and evolution of a full outburst.

\subsection{CASLEO optical observations}

Optical spectra of NSV 20407, the optical counterpart of \ax, were acquired using 
the 2.15-m 'Jorge Sahade' telescope at CASLEO (Argentina). This telescope is equipped 
with a REOSC spectrograph, which carries a 1024$\times$1024 pixel TEK CCD. 
The spectra were acquired in Simple Dispersion mode using the \#270 
grism (300 lines/mm) and a 2$''$ slit width, allowing one to nominally 
cover the 3500$-$7500 \AA~spectral range with a dispersion of 3.4 
\AA~pixel$^{-1}$.

Two 1200-s spectroscopic frames were secured on 2015 April 14, with start times of
04:54 and 05:16 UT, respectively. After cosmic ray rejection, the spectra were 
reduced, background subtracted and  extracted \citep{Horne86} using 
IRAF\footnote{Image Reduction and Analysis Facility (IRAF) is available at 
{\tt http://iraf.noao.edu/}} \citep{Tody93}. Wavelength calibration was performed 
using Cu-Ne-Ar lamps acquired before each spectroscopic exposure; the 
spectra were then
 flux-calibrated using the spectrophotometric standard LTT 6284
(Hamuy et al. 1994). Finally, the two spectra were stacked
 together to increase 
the signal-to-noise ratio. The wavelength calibration
 uncertainty was $\sim$0.5 \AA; 
this was checked using the positions of background night sky lines.

\subsection{Swift observations} 
\label{ob_swift}

We analysed all available observations since 2007 June 2 (including 
archival and new observations that we requested),  taken in Photon Counting 
(PC) mode with the X-Ray Telescope \citep[XRT; ][]{Burrows05} on board 
the \swift\ satellite.  The XRT observations with exposure times longer than 
500s were reduced using the \swift\ tools within the {\it heasoft} v. 6.16 \citep{Blackburn95}
software package.
Source detection and position determination were carried out using the recipes 
described in \cite{Evans09}.  The source light curves and spectra were extracted 
in the 0.5$-$10.0 keV band using a circular extraction region with a radius 
of 20 arcsec centred on the position of the source. Background data were extracted 
from an annular region with an inner (outer) radius of 30 arcsec (60 arcsec).
The 19 new \swift\ observations are shown in Table \ref{tab:xrt}.

Since \ax\ is not at the center source of the archival \swift\ observations and the UVOT 
field of view is smaller than XRT, the source was not observed in all UVOT 
observations. We reduced all available UVOT observations  which contained this 
source in the field. In each UVOT observation, this source was observed with at least one 
of the four filters with corresponding central wavelengths (Poole et al. 2008):
uvw2 (1928 \AA) uvm2 (2246 \AA), uvw1 (2600\AA ), and u (3465\AA). 
The UVOT data were analysed following the methods of Poole et al. (2008) and 
\cite{Brown09}.

\subsection{XMM-Newton observations}
\label{ob_xmm}

\begin{table*}
\begin{center}
\caption{The \xmm\ and \chan\ observations in which the source was detected. 
The source net count rates were calculated in the 0.3--10 keV energy band converted using {\sc pimms}.
Errors represent the 90\% confidence level.
We converted the \xmm\ and \chan\ count rate to equivalent \swift-XRT 
count rates. The last column shows the 3-$\sigma$  upper limit of the pulsed fraction
for each observation. 
}
\begin{tabular}{cccccccc}
\hline\hline
Instr. & No. & ObsID    & Date      & Exp. & Rate                & Rate(XRT/PC)    &  pulsed fraction upper limit \\
       &     &          &           & ks   & $10^{-2}~\cnts$     & $10^{-3}~\cnts$ &  \%  \\
\hline
\xmm\  & 1 & 0203910101 & 2004 Feb. & 9.1  & $11.3  \pm 0.5$     &  9.34 & 6.2 \\ 
\chan\ & 2 &  7287      & 2006 Jun  & 9.5  & $3.21  \pm 0.15$    &  4.71 & 6.5 \\
\xmm\  & 3 & 0402910101 & 2006 Aug. & 38.3 & $5.86  \pm 0.15$    &  4.91 & 6.1 \\
\xmm\  & 4 & 0410581901 & 2007 Aug. & 12.4 & $0.41  \pm 0.01$    &  0.39 & 6.6 \\
\xmm\  & 5 & 0560181101 & 2009 Feb. & 48.7 & $3.70  \pm 0.11$    &  3.11 & 6.5 \\
\xmm\  & 6 & 0604880101 & 2010 Feb. & 40.4 & $0.54  \pm 0.08$    &  0.52 & 5.4 \\
\chan\ & 7 &  12554     & 2011 Jun  & 96.5 & $0.78  \pm 0.03$    &  1.12 & 5.8 \\   
\hline
\end{tabular}
\label{tab:obs}
\end{center}
\end{table*}

We analysed five \xmm\ observations of the source between February 2004 and February 2010. 
We reduced the \xmm\ Observation Data Files (ODF) using version 12.0.1 of the 
Science Analysis Software ({\sc sas}). We used the {\sc epproc} task to extract 
the event files for the PN camera. 
Source light curves and spectra were extracted in the 0.5--10.0 keV band using a 
circular extraction region with a radius of 20 arcsec centred on the 
position of the source. Background light curves and spectra were extracted from a 
circular source-free region of radius 30 arcsec on the same CCD. We applied standard 
filtering and examined the light curves for background flares. Only 
Obsid 0402910101 contained flares and we used the non-flared exposures 
for our analysis. We checked the filtered event files for photon pile-up by 
running the task {\sc epatplot}. No pile-up was apparent in the data. Photon 
redistribution matrices and ancillary files were created using the {\sc sas} tools 
{\sc rmfgen} and {\sc arfgen}, respectively. We rebinned the source spectra 
using the tool {\sc grppha}, such that the minimum number of counts per bin of the 
PN spectra was 25\footnote{We found that there are no obvious differences between adding 
MOS data and using PN spectra only, therefore we only used PN data in this work.}. 

\subsection{Chandra observations}
\label{ob_chandra}

We analysed two \chan\ observations of \ax\ performed on Jun 2006 and Jun 2011,
respectively (Table \ref{tab:obs}).  We used the {\sc ciao} tools 
\citep[v. 4.5; ][]{Fruscione06} and standard \chan\ analysis threads to 
reduce the data. No background flares were found, so all data were 
used for further analysis.

The source spectra and light curves were extracted from a circular region with 
a radius of 20 arcsec  centred on the position of \ax. Background events 
were obtained from an annular region with an inner (outer) radius of 30 arcsec 
(60 arcsec). Using the {\sc ftool} {\sc grppha}, we re-binned the spectra to 
contain a minimum of 25 photons per bin.

\section{Analysis and Results}
\label{result}

\subsection{Optical light curves}
\label{optical}

\begin{figure}
    \centering
        \includegraphics[width=3.6in,angle=0]{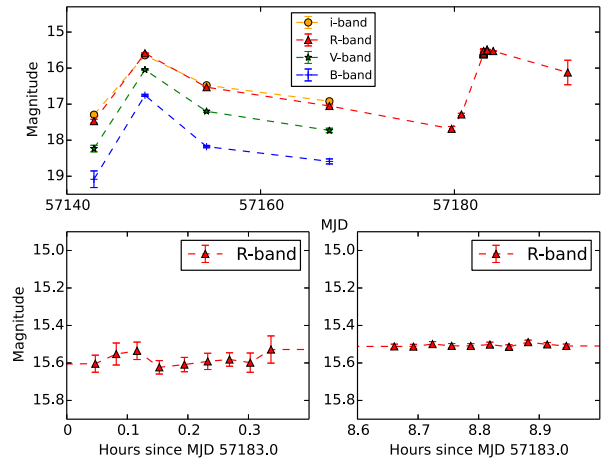}
    \caption{The optical light curve of the source in 2015. The top panel shows the 
    optical light curve of \ax\ in i$\arcmin$, R, V and B band with orange, red, green
    and blue colour, respectively. The bottom panels show two zoom-in plots on 
    the second peak with higher time resolution.}
    \label{fig lc optical}
\end{figure}

The top panel of Fig. \ref{fig lc optical} shows the optical light 
curve of \ax\ observed by LCOGT in April -- June 2015. The source showed 
two clear outbursts within $\sim 40$ days, 
with  the two peaks $\sim 35$ days apart. During the first outburst, the source was 
observed in $B$, $V$, $R$ and $i^{\prime}$-bands. During the second 
outburst, the source was observed in $R$-band only, with some higher 
time resolution ($\sim 2$ minutes) sequences on 2015 June 10. The 
lower panels of Fig. \ref{fig lc optical} show two zoom-in plots of 
the second peak. The source is less variable at short timescales (minutes)
than at longer timescales ($\sim$10 days).

The general morphology of the outbursts is consistent with a fast rise, exponential 
decay \citep[FRED; e.g.][]{Chen97}, with the optical spectrum becoming bluer 
at higher fluxes. The rise rates between the first and second dates were 2.3, 
2.2, 1.9 and 1.6 magnitudes in 5.26 days in $B$, $V$, $R$ and $i^{\prime}$-bands, 
respectively (0.44, 0.41, 0.36 and 0.31 mag/day). The rise rate of the second 
outburst was 2.2 mag in 3.3 days in $R$-band (0.65 mag/day), which may suggest 
the first outburst rise was quicker than measured since the first rise was
not well sampled. The decay rates between the second and fourth dates (over 
19.0 days) were 0.096, 0.088, 0.077 and 0.067 mag/day in $B$, $V$, $R$ and 
$i^{\prime}$-bands. The amplitude of the variations are larger at shorter 
wavelengths, which is expected for an accretion disc described by a simple 
blackbody that is hotter when it is brighter \citep[this is also the case 
for low-mass X-ray binaries; see][]{Maitra08, Russell11}.

Serendipitously, our monitoring caught the fast rise of the next
outburst, and the short time series coincided almost exactly with the peak 
of the outburst (Fig. \ref{fig lc optical}). On 2015-06-10, the average 
magnitude was $R = 15.58$ from data taken at 00:02--00:22 UT (9 consecutive
images, at a time resolution of 131 sec), and $R = 15.51$ at 08:39--08:58 UT.  
No intrinsic variability is detected; the magnitudes generally agreed within the
errors of each measurement. The conditions were good during the second time 
series and worse during the first, with mean magnitude measurement errors 
of 0.048 mag at 00:02--00:22 UT and 0.012 mag at 08:39--08:58 UT. The 
fractional rms variability  \citep[calculated adopting the prescription described 
in ][]{Gandhi10} was $< 13.7$ per cent at 00:02--00:22 UT and $< 3.2$ 
per cent at 08:39--08:58 UT ($3 \sigma$ upper limits).  To compare this 
result to observations of other cataclysmic variables, \cite{Sande15} report 
fractional rms variability intrinsic to the source in three CVs at a level 
of 1--5 per cent using data from the \emph{Kepler} satellite (the time 
resolution was 58.8 sec).

\begin{figure}
    \centering
        \includegraphics[width=3.5in,angle=0]{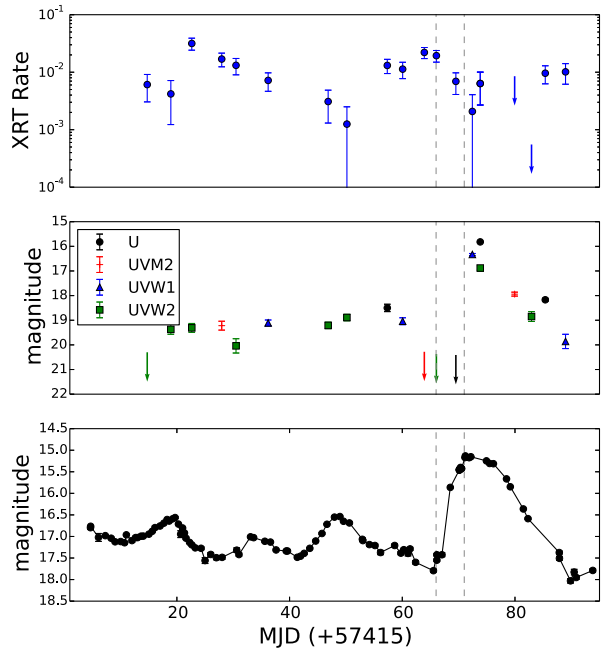}
    \caption{The  X-ray, UV and optical light curves of \ax\ are shown
    in the top, middle and bottom panels, respectively. The rising phase 
    of the outburst is indicated by two vertical dashed lines. The error 
    bars are given at the 1-$\sigma$ level. The observations that only have 
    upper limits are shown by arrows.}
    \label{fig lc optical 2016}
\end{figure}

In 2016 we continued monitoring the source in $R$-band only. 
We plot the R-band optical light curve of \ax\ with black filled circles in 
the bottom panel of Fig. \ref{fig lc optical 2016} between February and May 2016. 
Starting on Feb 1st 
(MJD 57419) the source showed three mini-outbursts within $\sim 45$ days. 
The time intervals between the three peaks are $\sim 15$ days.  
The rising and decaying times are comparable in these three mini-outbursts.
The amplitude of the variation is $\sim 1$ magnitude. 
Similar anomalous mini-outbursts have also been observed in the
DN system SS Cyg \citep{Schreiber03}.

After the three mini-outbursts the source again went into outburst, with the 
optical flux increased rapidly ( $\sim2.6$ mag 
within 5 days, $\sim$0.5 mag/day) and then decaying over the next $\sim15$ days,
both comparable to the two outbursts observed in 2015.
This outburst showed a clear FRED morphology and confirmed the DN nature.

In typical DN systems (e.g., SS Cyg, U Gem) the time interval of quiescence is generally
longer than the duration of an outburst. Whereas, in the 2015 and 2016 observations of
\ax, the duration at high flux levels is comparable to the time interval at low flux levels, 
and there appears to be no period of steady flux in quiescence. Instead, the  low
flux level periods are occupied by mini-outbursts and low level activity.

\subsection{Optical spectra}

\begin{figure}
    \centering
        \includegraphics[width=2.7in,angle=-90]{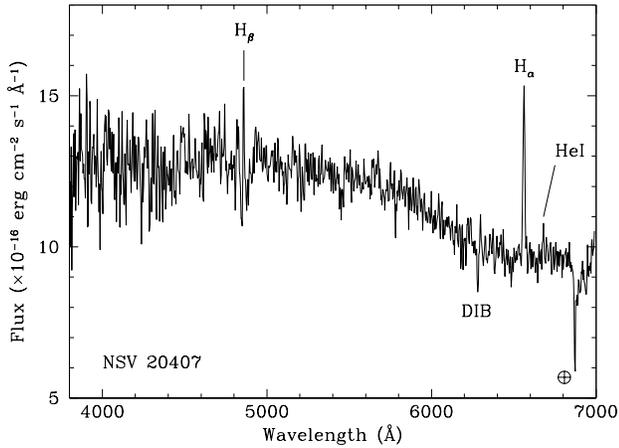}
    \caption{Averaged spectrum  of source NSV 20407 as observed at CASLEO on April 14, 2015 
    (see text for details). The main spectral features are shown. The symbol 
    $\oplus$ indicates atmospheric telluric absorption bands, while the 
    label 'DIB' marks the Galactic Diffuse Interstellar Band at 6280 \AA.}
    \label{fig: optical_pha}
\end{figure}

The average optical spectrum (not corrected for intervening Galactic 
absorption) of \ax\ (Fig. \ref{fig: optical_pha}) shows 
the presence of H$_\alpha$ and H$_\beta$ Balmer lines, and possibly He I $\lambda$6678,
in emission superimposed on an intrinsically blue continuum; the H$_\beta$ 
line seems to be embedded in the corresponding absorption feature.
All of the detected lines are consistent with being at a redshift of $z$ = 0, indicating 
that this object is within our Galaxy.
Fluxes and equivalent widths (EWs) of the two Balmer lines are: 
$F_{\rm H_\alpha}$ = (8.5$\pm$1.5)$\times$10$^{-15}$ erg cm$^{-2}$ s$^{-1}$;
$F_{\rm H_\beta}$ = (3$\pm$1)$\times$10$^{-15}$ erg cm$^{-2}$ s$^{-1}$;
$EW_{\rm H_\alpha}$ = 9.1$\pm$1.6 \AA; $EW_{\rm H_\beta}$ = 3$\pm$1 \AA.
No He II emission at 4686 \AA~is detected down to flux and EW 
3-$\sigma$ limits of 2$\times10^{-15}$ erg cm$^{-2}$ s$^{-1}$ and 2 \AA, 
respectively. These properties do not vary significantly between the two
spectra.

Despite the non-optimal signal-to-noise ratio of the optical spectrum,
especially in its blue part, its main characteristics depicted above support
the identification of NSV 20407 as a CV. Moreover, the lack of apparent
He II emission suggests a non-magnetic nature of the accreting WD
in this system \citep[see][and references therein for details]{Warner95}. 
Also, the observed H$_\alpha$/H$_\beta$ flux ratio ($\sim$2.8), when compared
with the intrinsic one \citep[2.86; ][]{Osterbrock89} is  typical of plasmas in 
accretion in astrophysical conditions, and points to no substantial reddening
in the line of sight towards the object.

\subsection{X-ray light curve}
\label{swift lc}

\begin{figure*}
    \centering
    \includegraphics[width=7in,angle=0]{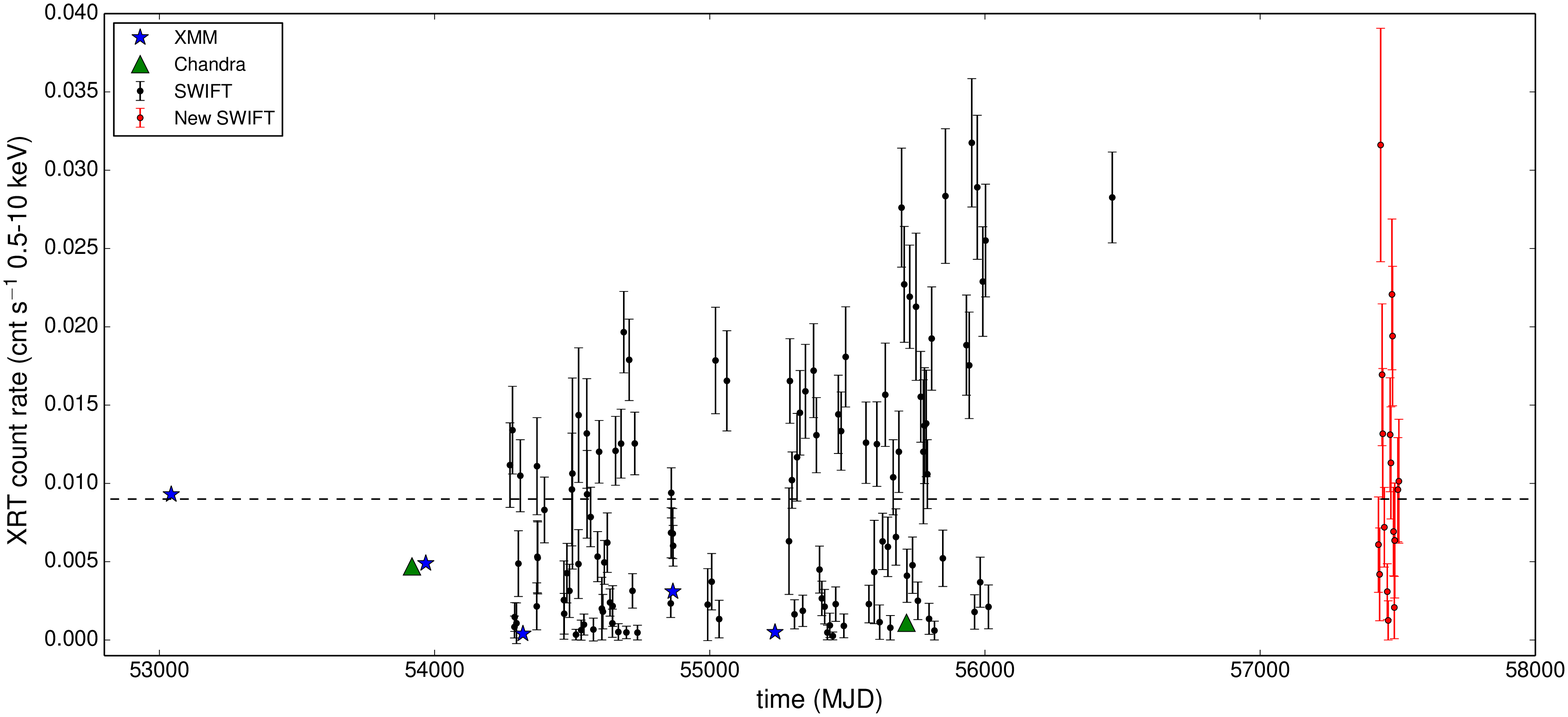}
    \caption{ The \swift/XRT light curve of AX J1549.8--5416 (one bin per 
    observation) is represented by black and red (2016 observations) filled circles. The error bars are 
    given at the 1-$\sigma$ level. The blue stars indicate the \xmm\ EPIC-pn 
    count rate and the green triangles are the \chan\ count rate, respectively (converted to the 
    equivalent XRT count rate). The typical spacing between \swift\ 
    observations is $\sim 1-2$ weeks in the data before 2016. 
    The horizontal dashed line indicates the 0.009 cnt/s level below which the 
    source is in optical outburst. 
    }
    \label{fig xrt lc}
\end{figure*}

Fig. \ref{fig xrt lc} shows the long-term X-ray light curve of \ax\ observed 
by \swift, which indicates its  X-ray flux varies by two orders of magnitude.
Possibly due to  observations being separated by large, irregular amounts of time,
we do not find any clear trend in the X-ray light curve.

\begin{figure}
    \centering
        \includegraphics[width=3.4in,angle=0]{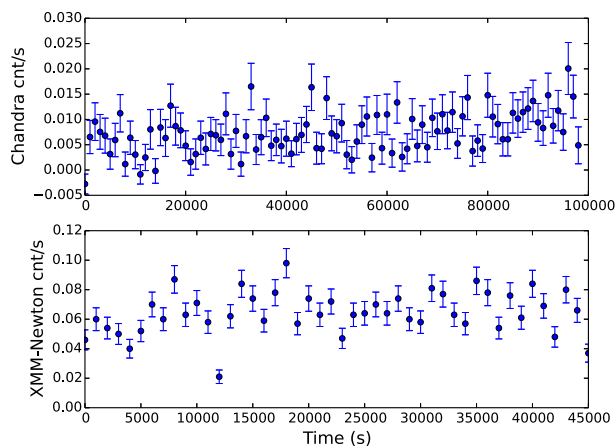}
    \caption{Top panel: the light curve of \ax\ observed by \chan\ in Obsid 12554 in Jun 2011;
    bottom panel: the light curve of the source observed by \xmm\ in Obsid  0402910101 in Aug 2006. 
    Both light curves are binned at 1000s. }
    \label{fig xmm_chan_lc}
\end{figure}

We converted the \xmm\ and \chan\ count rates to equivalent \swift-XRT count rates 
using {\sc pimms} in {\sc ftools} based on the fitted  power law indexes 
shown in Table \ref{tab:fit}, with results given in Table \ref{tab:obs} and 
shown in Figure \ref{fig xrt lc}, respectively.
The longer and more sensitive \chan\ and \xmm\ observations also allowed us  
to search for significant changes in the X-ray flux on short time-scales.
Fig. \ref{fig xmm_chan_lc} shows two light curve examples observed by \chan\ and \xmm.
At a time resolution of 1000 s, the X-ray intensity varied between 0.001 and 0.015 counts/s, 
and 0.02 and 0.1  counts/s in \chan\ and \xmm\ observations, 
but no clear outbursts are observed.

We also searched for periodic X-ray variability in these observations, 
which could correspond to the spin or orbital period of the WD.
To search for a periodic modulation in the X-ray light curves we created the Leahy power
density spectra (PDS) \citep{Leahy83} for each observation. We used the pn and ACIS
light curves extracted from the source region and binned at the frame readout time from
\xmm\ and  \chan\, respectively. We searched for a periodic behaviour of the 
source in a range between 73.4 ms and 26.8 hour.  No significant periodicity 
(0.5--10 keV) was found in all \xmm\ and \chan\ observations. 
We show the 3-$\sigma$ upper limit of the pulsed fraction for 
each observation in Table \ref{tab:obs}. {The lack of a coherent signal at the 
WD spin period in the X-ray band suggests that the WD does not possess a strong 
magnetic field. }

\subsection{Multi-wavelength observations in 2016}
\label{}

Fig. \ref{fig lc optical 2016} shows the X-ray/UV/optical observations of 
\ax\ from February to May 2016.  During the phase when the source showed three
mini-outbursts, we do not find a clear trend in the UV bands, probably because of the changes
between the UVOT filters. At X-ray energies the source shows $\sim 2$ orders of magnitude 
variations during these three mini-outbursts. The X-ray flux is highest when the 
source is in optical quiescence. After the three mini-outbursts, both the optical 
and UV fluxes decreased and the source went to the quiescent state. The UV emission was not 
detected before the largest optical outburst.

In the rising phase of the largest optical outburst, the UV emission was 
still undetected, and the X-ray flux decreased. The UV emission appeared at the 
time of optical peak, which indicates a UV delay in the rising phase.

Both optical and UV fluxes decreased during the early decaying phase. The X-ray 
emission was not detectable in this phase. 
At the end of the outburst the optical and UV fluxes 
decreased continually, but the X-ray emission recovered.  Both the X-ray 
suppression during the outburst and recovery at the end of the outburst have 
also been reported in another DN, SS Cyg \citep{Wheatley03}.

\subsection{X-ray $-$ UV and X-ray $-$ optical correlation}
\label{x-ray and uv}

\begin{figure}
    \centering
    \includegraphics[width=3.3in,angle=0]{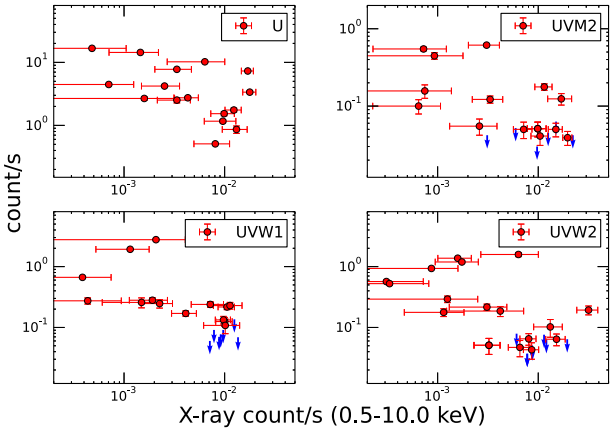}
    \caption{Simultaneous UV$-$X-ray intensity diagrams for four UVOT filters. 
    The error bars are given at the 1-$\sigma$ level. The blue arrows indicate UV upper
    limits.}
    \label{fig xray and uv}
\end{figure}

Fig. \ref{fig xray and uv} shows  the simultaneous UV$-$X-ray intensity 
correlation diagrams of \ax\ for the four UVOT filters.  
In the four UV bands, the X-ray flux is to some extend anti-correlated 
with the UV flux in this source. In order to describe the correlation 
quantitatively, we used a constant function and a constant plus a power law function to fit 
the data in four UVOT filters, respectively. The fitting results are shown in 
Table \ref{tab:fit con powerlaw}. The f-test probability suggests that,  
instead of using a constant function, the data can be better fitted by adding
a power law function. The power law indices are all negative and indicate
that the UV and X-ray are anti-correlated  in the four filters. 
We also performed a Spearman's rank correlation coefficient test between X-ray and UV fluxes. 
and report in Table \ref{tab:fit con powerlaw} the value  of the
Spearman's rank correlation coefficient and the null hypothesis probability (p-value). 
The negative correlation coefficient and small p-values  again 
indicate that the UV and X-ray are anti-correlated.  We note that the observations with 
upper limits are not used in the above analysis.

\begin{table*}
\begin{center}
\caption{The fitting parameters of a constant function,  a power law with a constant function, 
correlation coefficient and p-values of Spearman's rank correlation coefficient
test for the UV $-$ X-ray correlation.  
}
\begin{tabular}{ccccccccccccc}
\hline\hline
data   & constant $\chi^{2}$ (dof) & $\Gamma$       & power law $+$ constant  $\chi^{2}$ (dof) & f-test probably    &  correlation coefficient & p-value ($\%$) \\
\hline
U     &  159.7(15)                  & $-2.2\pm 0.8$    & 89.5(13)                   &  $2.3\times 10^{-2}$      &  -0.52                  & 3.87    \\  
UVM2  &  217(14)                    & $-3.8\pm 1.1$    & 58.3(12)                   &  $3.7\times 10^{-4}$      &  -0.54                  & 3.59    \\
UVW1  &  156(15)                    & $-0.6\pm 0.2$    & 29.4(13)                   &  $1.9\times 10^{-5}$      &  -0.76                  & 0.03    \\
UVW2  &  120(17)                    & $-0.2\pm 0.1$    & 65.5(15)                   &  $3.9\times 10^{-3}$      &  -0.74                  & 2.03    \\
\hline
\end{tabular}
\label{tab:fit con powerlaw}
\end{center}
\end{table*}

\begin{figure}
    \centering
        \includegraphics[width=2.6in,angle=0]{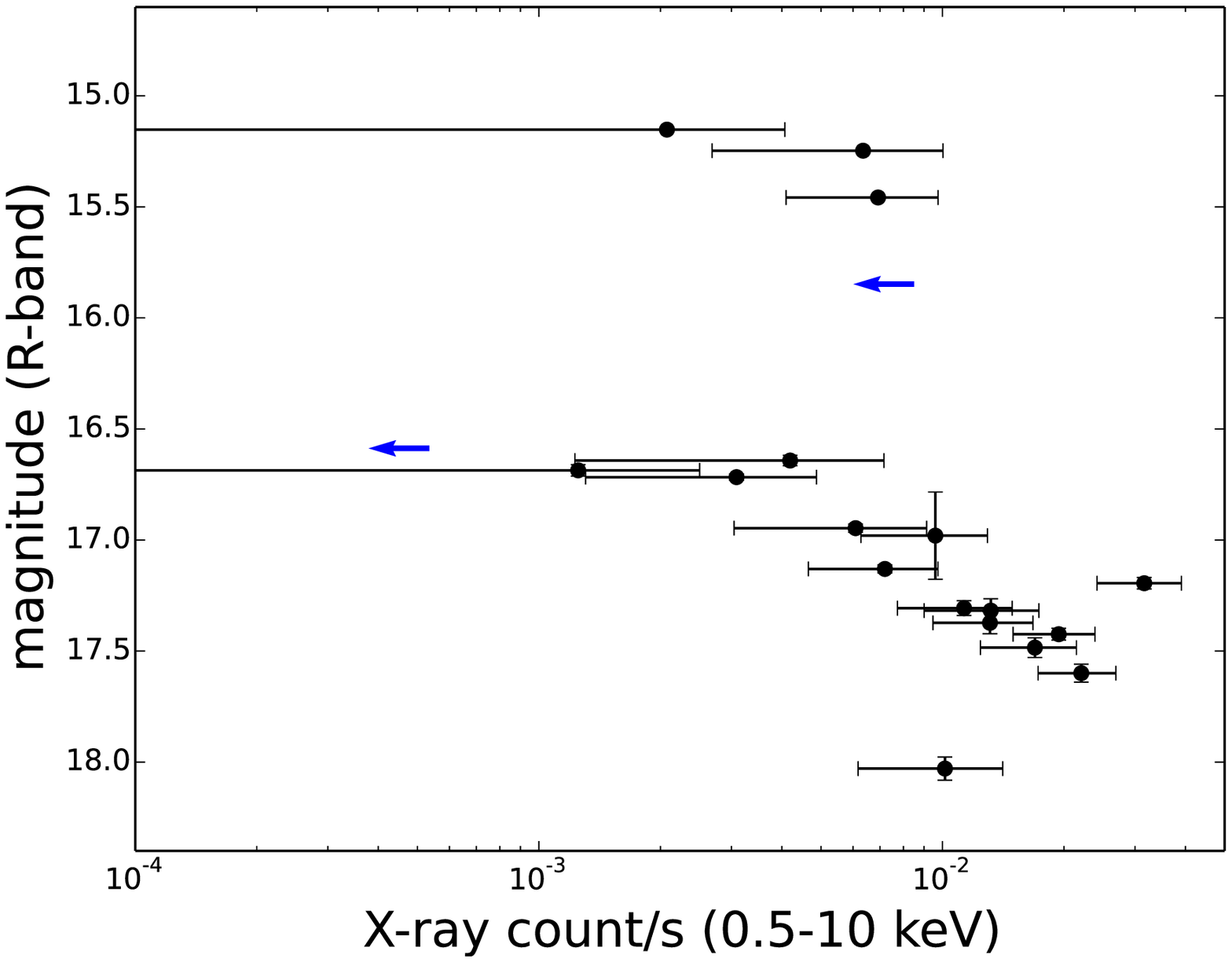}
    \caption{Simultaneous X-ray$-$ optical flux correlation of \ax. 
    The error bars are given at the 1-$\sigma$ level. The blue arrows indicate 
    X-ray the upper  limits. }
    \label{fig xray and optical}
\end{figure}

Fig. \ref{fig xray and optical} shows the simultaneous optical$-$X-ray flux 
correlation of \ax\ during the 2016 observations. We found that the X-ray and 
optical flux also show an anti-correlation, which is much clearer at low 
optical fluxes.

\subsection{X-ray Spectral analysis}
\label{X-ray Spectral}

\begin{figure}
    \centering
    \includegraphics[width=3.7in,angle=0]{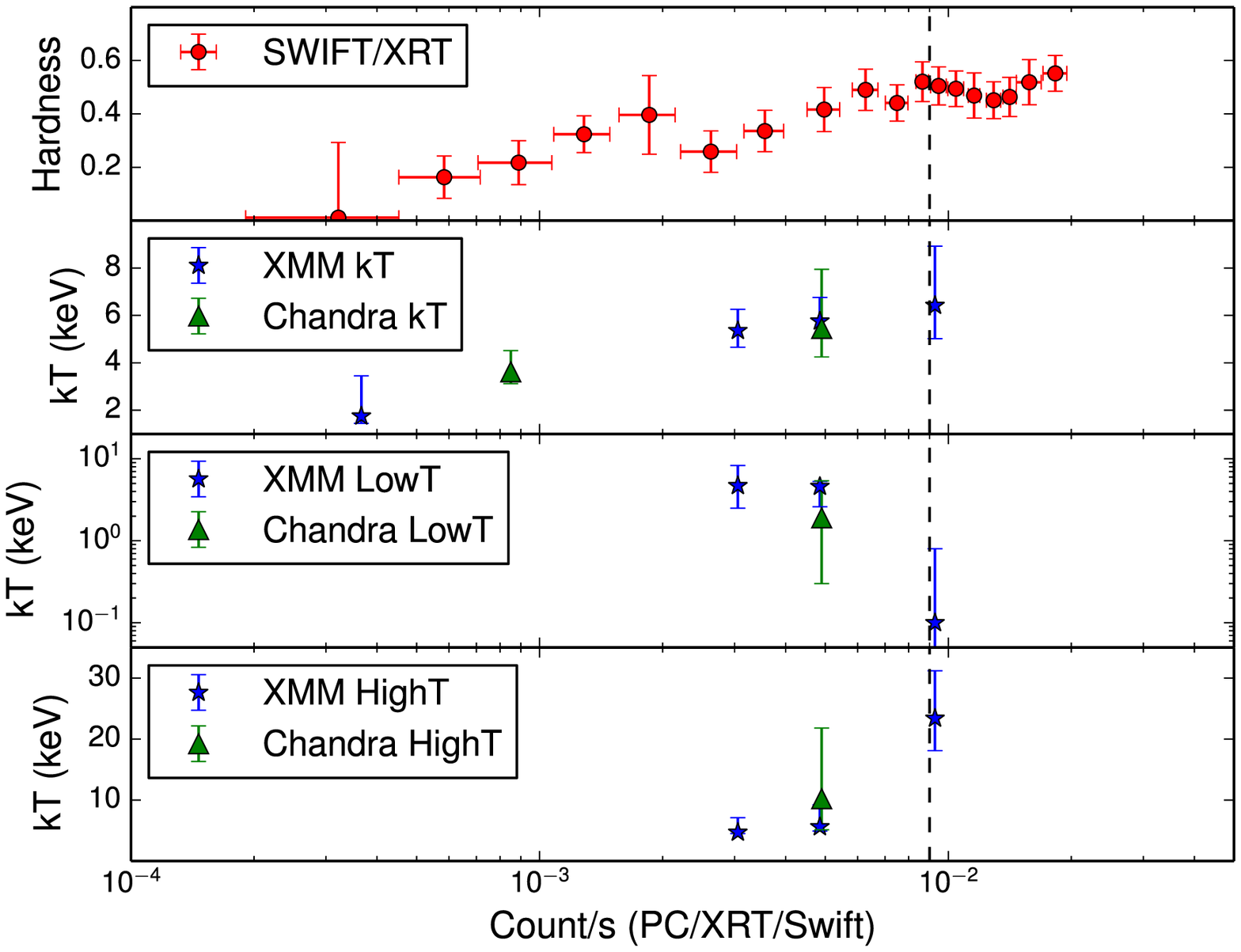}
    \caption{Top panel: the hardness-intensity diagram of AX J1549.8--5416 observed by \swift.
     Second panel: the plasma temperature (in {\sc mekal}) as a function of X-ray intensity. 
     Third and fourth panels: the kT$_{Low}$ and kT$_{High}$ (in {\sc mkcflow}) as a function
     of X-ray intensity. We converted the \xmm\ and \chan\ intensity to equivalent 
    \swift\  intensity. The vertical dashed line indicates the 0.009 cnt/s level. 
    } 
    \label{fig swift_xmm_chan_hrd}
\end{figure}

Due to the low signal-to-noise ratio of this source in an individual \swift\ 
XRT observation, spectral fitting is not possible, so we used the X-ray colour 
to study the X-ray spectra with \swift.  
For each observation, we calculated the X-ray hardness. We defined the hardness 
as the count rate in the 2.0--10.0 keV band divided by the count rate in 
the 0.3--10.0 keV band. In the upper panel of Fig. \ref{fig swift_xmm_chan_hrd}
we show the hardness-intensity diagram (HID) of the source.  The data are 
rebinned so that each bin has approximately the same counts. As shown in the upper 
panel of Fig. \ref{fig swift_xmm_chan_hrd}, 
the X-ray spectrum changes with its flux. 
Above $\sim 2\times 10^{-3}$ cnt/s, the X-ray emission is hard, 
with a constant hardness ratio  $\sim$ 0.6.  Below this flux level, its X-ray 
emission becomes soft. We first fitted these data with a constant function 
and obtained a $\chi^2$ of 151.2 
(dof = 17). We then fitted these data with a power-law model, and got a power-law 
index of $0.57 \pm 0.07$ (1-$\sigma$ error) with a $\chi^2$ of 41.2 (dof = 16). 
The smaller reduced $\chi^2$ indicates that a power-law fit is better than a 
constant. We performed a Kolmogorov-Smirnov (K--S) test of our count rates sample 
against the standard normal distribution. The probability value of K$-$S test 
is 0.12, which indicates the source count rates are not consistent with random 
points around a constant value.

We also measured the X-ray spectrum of the source from four \xmm\ and two \chan\ 
X-ray observations (observation 0410581901 was not used due to the low count 
rate and high background). The spectra were fitted in 0.5$-$10 keV range with 
{\sc xspec} \citep[v. 12.8.1][]{Arnaud96}. We included the effect of interstellar 
absorption using {\sc wabs} assuming cross-sections of \cite{Balucinska92} and 
solar abundances from \cite{Anders89}, and we let $N_{\rm H}$, the column 
density along the line of sight, vary during the fitting.

In order to understand the source spectral shape at different flux levels, 
we first used a power-law model to fit all spectra. 
The continuum spectra can be well fitted by a power-law model except some
residuals around 6.7 keV, which may come from Iron lines. The $N_{\rm H}$
shows comparable values, $\sim (4-5) \times 10^{21}$ cm$^{-2}$, in all spectra. The fitted
power-law parameters are shown in Table \ref{tab:fit}.
The spectrum photon index is larger when the source is at a low X-ray flux level
than at a high X-ray flux level. This is consistent with the XRT X-ray hardness 
analysis above. There is no need to include a soft component below 1 keV in all
spectral fits.

\begin{table*}
\begin{center}
\caption{Fit statistics and parameters of the spectra models {\sc powerlaw} 
and {\sc mekal} applied to the six X-ray spectrum of AX J1549.8--5416. 
Error bars are given in 90\% confidence levels.}
\begin{tabular}{ccccccccccccc}
\hline\hline
ObsID        & $N_{\rm H}$ (PL)       & $\Gamma_{\rm PL}$      &  reduced $\chi^{2}$ (PL) & $N_{\rm H}$ (MK)        &    kT    & norm &reduced $\chi^{2}$ (MK)  \\
             & $10^{22}$ cm$^{2}$     &                        &                  & $10^{22}$ cm$^{2}$      &    keV                 &      $\times 10^{-4}$            \\
\hline
 0203910101  & $0.38^{+0.11}_{-0.10}$ & $1.75^{+0.22}_{-0.21}$ & 1.01             & $0.31^{+0.08}_{-0.07}$  &  $6.81^{+3.41}_{-1.84}$ & $7.4^{+0.5}_{-0.7}$ & 0.91            \\ 
 7287        & $0.36^{+0.27}_{-0.23}$ & $1.85^{+0.42}_{-0.38}$ & 0.94             & $0.33^{+0.21}_{-0.18}$  &  $5.12^{+5.62}_{-1.69}$ & $3.4^{+0.7}_{-0.5}$ &0.87            \\
 0402910101  & $0.50^{+0.07}_{-0.06}$ & $2.04^{+0.14}_{-0.13}$ & 1.71             & $0.35^{+0.05}_{-0.05}$  &  $5.45^{+1.00}_{-0.79}$ & $3.5^{+0.2}_{-0.2}$ &1.21            \\
 0560181101  & $0.49^{+0.07}_{-0.06}$ & $2.05^{+0.16}_{-0.15}$ & 1.37             & $0.36^{+0.05}_{-0.04}$  &  $5.02^{+0.96}_{-0.75}$ & $2.9^{+0.2}_{-0.2}$ &1.32            \\
 0604880101  & $0.52^{+0.33}_{-0.26}$ & $3.49^{+1.36}_{-0.91}$ & 2.21             & $0.04^{+0.15}_{-0.04}$  &  $2.67^{+1.22}_{-0.71}$ & $0.35^{+0.10}_{-0.07}$&1.57            \\
 12554       & $0.45^{+0.34}_{-0.21}$ & $2.45^{+0.21}_{-0.19}$ & 1.21             & $0.05^{+0.07}_{-0.04}$  &  $6.08^{+3.01}_{-1.45}$ & $0.77^{+0.07}_{-0.07}$&0.91            \\   
\hline
\end{tabular}
\label{tab:fit}
\end{center}
\end{table*}

As argued above, the optical light curves and spectra indicate the source is a DN.
The simple model which is often used in DNe consists of a single-temperature optically 
thin thermal plasma model \citep[{\sc mekal} in {\sc XSPEC}, e.g. ][]{Mewe85}. 
This model  has been successfully used in $\sim$30  spectra  of dwarf novae  
\citep{Baskill05} observed with ASCA. We also used this model to fit all our X-ray 
spectra separately. Because the abundance derived by {\sc mekal} is consistent with  solar abundance 
within uncertainty, we then fixed the abundance to 1 in our analysis. 
The fitted parameters are shown in Table \ref{tab:fit}. With the same 
degree of freedom and smaller $\chi^2$ in each observation, the {\sc mekal} model 
gives a better fit than the power-law model.

\begin{figure}
    \centering
        \includegraphics[width=2.35in,angle=-90]{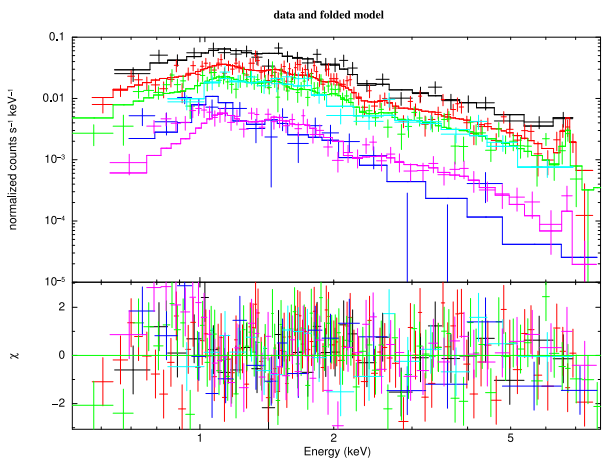}
    \caption{ The \xmm\ PN, and \chan\ ACIS spectra of AX J1549.8--5416 
    in the 0.5$-$10.0 keV energy band. The spectrum was fitted with a 
    single-temperature thin thermal plasma model. The six spectra are represented
    by six different colour  curves in the upper panel.
    The lower panel shows the residuals to the best-fitting models. 
    The observations 1, 2, 3, 5, 6 and 7 (in Table-\ref{tab:obs}) 
    are shown in black, light blue, red, green, purple and blue, 
    respectively. }
    \label{fig xmm_chan_spectra}
\end{figure}

From Table \ref{tab:fit} the $N_{\rm H}$ shows a similar value at different
X-ray flux levels. In order to better understand the evolution of the thermal plasma 
temperature at different flux levels, we fitted the six spectra 
simultaneously with linked $N_{\rm H}$. The $N_{\rm H}$ value from the 
simultaneous fitting is ($0.33\pm 0.03$) $\times 10^{22}$ cm$^{2}$, 
which is comparable with values from individual spectrum fits in Table \ref{tab:fit_all}. 
Fig. \ref{fig xmm_chan_spectra} shows the X-ray spectrum of each observation 
(four \xmm\ and two \chan) overlaid with best-fitting {\sc mekal} model.  
The best-fitting reduced $\chi^{2}$ is  1.29 with 296 degrees of freedom. The 
iron line is clearly detected in {bf the} four observations which have higher X-ray flux.

\begin{table*}
\begin{center}
\caption{Fit results for the  {\sc mekal} model  applied 
to six combined X-ray spectra of AX J1549.8--5416. The absorbed 0.5$-$10 keV flux are also reported. 
Error bars are given at the 90\% confidence level. The $N_{\rm H}$ was 
linked across the spectra and the value is $0.33\pm 0.03 \times 10^{22}$ cm$^{2}$.  }
\begin{tabular}{cccccccccccccc}
\hline\hline
ObsID        &      kT                  & Normalization (MK)             & flux (0.5$-$10 keV)         \\
             &      keV                 &    $10^{-4}$                   &  $10^{-13}$ erg/cm$^2$/s    \\
\hline
 0203910101  &    $6.42^{+2.53}_{-1.43}$ & $7.56^{+0.56}_{-0.55}$        &    $11.76^{+1.23}_{-2.05}$  \\ 
 7287        &    $5.15^{+2.52}_{-1.23}$ & $3.47^{+0.35}_{-0.34}$        &    $5.01^{+0.71}_{-0.51}$   \\
 0402910101  &    $5.76^{+1.03}_{-0.73}$ & $3.41^{+0.17}_{-0.16}$        &    $5.13^{+0.23}_{-0.21}$   \\
 0560181101  &    $5.36^{+0.98}_{-0.78}$ & $2.84^{+0.15}_{-0.15}$        &    $4.15^{+0.19}_{-0.41}$   \\
 0604880101  &    $1.75^{+0.72}_{-0.38}$ & $0.43^{+0.13}_{-0.09}$        &    $0.43^{+0.11}_{-0.12}$   \\
 12554       &    $3.62^{+0.93}_{-0.56}$ & $0.96^{+0.07}_{-0.07}$        &    $1.18^{+0.05}_{-0.09}$   \\   
\hline
\end{tabular}
\label{tab:fit_all}
\end{center}
\end{table*}

It is clear from Table \ref{tab:fit_all} that the temperature of the thermal plasma 
is higher when the source has higher X-ray flux. The second panel of Fig.-\ref{fig swift_xmm_chan_hrd} 
shows the plasma temperature as a function of X-ray  
intensity. When the source is in the high intensity state, 
the X-ray spectra show comparable plasma temperature at $\sim 6$ keV. When the 
source evolves to low intensity levels, the plasma temperature decreases 
significantly to $\sim$ 1.5 keV. We find that the normalization of {\sc mekal} 
also decreases as the X-ray flux decreases.

As discussed in Section \ref{x-ray and uv}, the large variations in both X-ray 
and UV bands and the anti-correlation between X-ray and UV emission suggest that 
the source is a DN \citep{Wheatley03, Collins10, Britt15}. We therefore also
used an isobaric cooling flow model, {\sc mkcflow}, to fit the X-ray 
emission from the source \citep{Mushotzky88}. This model is often used to describe the X-ray 
spectrum of DNe in optical quiescence \citep{Pandel05,Mukai09, Byckling10}. 
Unlike the single temperature  plasma model, the cooling flow model is assumed 
to consist of a range of temperatures. In the model, the temperature varies 
from the hot shock temperature (kT$_{High}$) to the temperature of the optically 
thin cooling matter (kT$_{Low}$) on the WD surface. Thus, cooling flow spectral 
models should represent a more physically correct picture of the cooling plasma.

\begin{table*}
\begin{center}
\caption{Fit parameters of the spectral models {\sc mkcflow}  applied 
to the four X-ray spectra of AX J1549.8--5416.  Error bars are given at the 90\% confidence level.}
\begin{tabular}{cccccccccccccc}
\hline\hline
ObsID        & kT$_{Low}$           &   kT$_{High}$          & Normalization (MK)        & flux (0.5$-$10 keV)   \\
             & keV                  &    keV                 &    $10^{-11} \Msun$ /yr     & $10^{-13}$ erg/cm$^2$/s \\
\hline
 0203910101  & $0.1^{+1.4}_{-0.08}$ & $23.5^{+7.8}_{-5.3}$   & $10.7^{+3.4}_{-3.3}$      & $11.2^{+1.5}_{-3.2}$ \\ 
 7287        & $1.9^{+4.8}_{-1.8}$  & $10.5^{+20.5}_{-6.1}$  & $10.8^{+290.7}_{-7.4}$    & $4.8^{+0.2}_{-1.2}$\\
 0402910101  & $4.1^{+0.6}_{-2.6}$  & $6.3^{+6.2}_{-1.5}$    & $65.6^{+223}_{-45.4}$     & $4.9^{+0.2}_{-02}$\\
 0560181101  & $3.8^{+1.4}_{-2.1}$  & $5.8^{+7.6}_{-1.5}$    & $120.5^{+354.8}_{-95.4}$  & $4.0^{+0.1}_{-0.4}$\\
\hline
\end{tabular}
\label{tab:fit_all_cflow}
\end{center}
\end{table*}

Due to the low X-ray count rate and the numerous parameters in {\sc mkcflow}, we only 
used this model to fit the spectrum obtained from observations with $> 300$ counts.
In the following analysis we only used observations 1, 2, 4 and 5 (see Table \ref{tab:obs}).
The parameters are shown in Table \ref{tab:fit_all_cflow}. We note that, as the X-ray
flux decreases, the normalization (accretion rate) increases, kT$_{Low}$ increases and
kT$_{High}$ decreases. The temperature separation (kT$_{High}$ -- kT$_{Low}$ ) decreases
when the X-ray flux decreases.

Further-more, we first linked the kT$_{High}$ to the same value in 
all four observations and then let it be free in observation 1. 
The  $\chi^{2}$ reduced from 278.15 (227 dof) to 268.21 (226 dof). 
This indicates that the source in observations 2, 4 and 5 has a different 
X-ray spectrum than that in observation 1.

\section{Discussion}
\label{discussion}

\subsection{\ax\ is a DN}
\label{sec DN}

We have analysed new optical light curves and spectra of the  
CV \ax. We detected three outbursts and three mini-outbursts in the 2015 and 2016
optical light curve. We also detected H$_\alpha$ and H$_\beta$ lines in the 
optical spectrum.  Both the fast rise and exponential decay optical outbursts 
and the lack of apparent He II emission lines in the optical spectra 
suggest that \ax\ is a non-magnetic CV. This is not consistent with 
the MCV classification reported by \cite{Lin14}.

We then analysed all available archival data from \swift,  \xmm\ and \chan,
and find that the source is also highly variable at UV and X-ray wavelengths. 
We find that the X-ray intensity increases with X-ray hardness.   
We also find an anti-correlation between X-ray 
and UV flux in this source. The high-resolution X-ray spectra of \xmm\ and 
\chan\ can be well described either by a single temperature thermal plasma 
model or by an isobaric cooling flow model when its X-ray flux is high.
The behavior of the X-ray  and UV evolution in this system is very similar 
to other DNe \citep[e.g. SS Cyg, SU UMa, ][]{Wheatley03, McGowan04, Kording08}, 
and further confirms that \ax\ is a DN.

From X-ray timing analysis of this system we did not find a significant periodic 
signal from all \xmm\ and \chan\ observations, which is in common 
with other DNe. \cite{Baskill05} analyzed 34 non-MCVs (20 are DNe) observed 
by {\it ASCA} and found that most of the sources do not show periodic modulation 
of their X-ray flux.

\subsection{UV delay and X-ray suppression during the outburst in \ax}
\label{sec X-ray/uv delay}

We studied the behaviour of the optical, UV and X-ray emission 
during the outburst to help us to understand the movement of material 
through the accretion disc. We observed \ax\ in 2016 with a dedicated optical, 
UV and X-ray campaign and detected three
mini-outbursts and one normal outburst. We find that the UV emission is clearly 
delayed  with respect to the optical emission during the rising phase of 
the outburst (see Fig. \ref{fig lc optical 2016}). 
The first optical rise occurred between MJD 57482.04 and 57483.52.
The UV emission was still not detectable on MJD 57484.51, so the UV delay 
must be longer than 1.0 day. The UV became bright on MJD 574487.44, which indicates
the UV delay must be less than 5.4 days.  Due to the long separation between 
\swift\ observations, we can therefore only constrain the  delay time to be $\sim 1.0-5.4$ day,
The UV delay during the rising phase of DN outbursts has already been reported
\citep[e.g., VW Hydri, U Gem and SS Cyg, ][]{Cannizzo01, Wheatley03}.
\cite{Wheatley03} find that in DN SS Cyg the UV emission delays the optical 
emission by $\sim 1.5-2$ days during the outburst rise, comparable to 
what we infer for \ax.

\cite{Wheatley03} find that in SS Cyg the X-ray flux first rises fast and then drops 
immediately during the optical outburst rising phase \citep[see also ][]{Russell16}.  
During the rise of the outburst in \ax, we also detect X-ray emission significantly lower 
than that observed  before the outburst (quiescence). However, we do not find a spike  
in X-ray during the outburst rise of \ax. Since the separation of our \swift\ 
observations is $\sim$ 3 days and the duration of the X-ray spike in SS Cyg 
is $\sim$ 1 day, it is likely that we missed the spike shape variation in our X-ray observations. 
Similar to SS Cyg we also find the suppression and recovery of the X-ray emission
during the middle and the end of the optical outburst, respectively.

Outbursts of DNe are generally interpreted as a sudden brightening of 
the optically thick accretion disc due to an increased accretion rate 
\citep[e.g., ][]{Osaki96, Lasota01}. The UV delay time has been used 
to measure the heating wave propagation time \citep{Cannizzo01, Wheatley03, Schreiber03}.
Unfortunately our optical/UV data are not sufficient to allow us to
estimate the beginning and timescale of the propagation of the heating wave.

The suppression of the X-ray emission during an outburst has been interpreted as 
the boundary layer becoming optically thick to its own radiation. 
At the end of the outburst the boundary layer switches back to optically thin, 
the inner disc truncates and emits in the X-ray  band \citep{Cannizzo01, Wheatley03}.

\subsection{Mini-outbursts}
\label{sec mini-outburst}

\ax\ showed three mini-outbursts in the 2016 optical observations. 
Similar anomalous outbursts have also been observed in SS Cyg \citep{Schreiber03}. 
These kind of mini-outbursts cannot be reproduced by the standard 
disc instability model (DIM). \cite{Schreiber03} managed to reproduce
the mini-outbursts by modifying the DIM.  The simulated mini-outburst 
light curve shows a short rise time and a sharp peak. However, the observed
mini-outbursts show rise times ($\sim 9$ days) and similarly slow decays. 
In their simulations,  they assume the disc extends
down to the surface of the WD (no truncation) and heating of the outer 
disc and tidal dissipation are completely neglected.

\subsection{X-ray spectra of \ax\ in outburst and quiescence}
\label{sec X-ray spectra}

The X-ray spectra of DNe changes dramatically between optical quiescence and 
outburst. In quiescence,  X-ray spectra are normally described by multi-temperature
optically thin thermal models, such as observed from cooling flows. In outburst, 
the emission is instead described by optically thick emission in the EUV band, 
with characteristic temperatures around 10 eV. The X-ray spectrum is softer 
and the temperature is lower than that in optical quiescence \citep[e.g., ][]{Baskill05, Collins10}. 
In Fig. \ref{fig swift_xmm_chan_hrd}, we show that the plasma temperature  
decreases as the X-ray flux decreases (optical flux increases).  
The evolution of X-ray spectra of  \ax\ can be well interpreted by the DN hypothesis.

We also used an isobaric cooling flow model to re-analyse the four observations 
(Obs 1, 2, 3 and 5) when the source had higher X-ray flux. This model is 
often used to describe the  quiescent X-ray spectrum of DNe \citep{Pandel05,Mukai09, Byckling10}. 
The four spectra are well fitted by the cooling flow model.   
As the X-ray flux decreases, the temperature separation (kT$_{High}$ -- kT$_{Low}$) decreases
as well, which indicates the X-ray emission became optically thick.
This is also consistent with the DN hypothesis \citep{Kuulkers06, Patterson11, Balman12}.

\cite{Ishida09} analysed data from {\it Suzaku} observations of the DN SS Cyg in 
quiescence and outburst. They found the maximum temperature of the plasma is $\sim 20$ keV
and $\sim 6$ keV  in quiescence and outburst, respectively. \cite{McGowan04} 
also found that the plasma temperature is $\sim$20 keV when SS Cyg is in quiescence. 
From Table \ref{tab:fit_all_cflow} we find that only Obs 1 has a maximum plasma
temperature higher than 20 keV. This suggests that only Obs 1 was observed in 
quiescence if we assume \ax\ and SS Cyg have comparable maximum plasma temperature 
in quiescence and outburst. The maximum temperature of the plasma in quiescence
is significantly higher than that in outburst in \ax.

\subsection{Duty cycle}
\label{duty_cycle}

In DNe the X-ray flux is always suppressed during optical outbursts and then 
shows a general anti-correlation with optical flux \citep{Wheatley03, McGowan04}.
Since we have a large number of XRT observations, instead of using optical observations, we
can use the X-ray data to roughly estimate the duty cycle of \ax. 
We define the outburst state when the X-ray intensity is lower than 0.009 cnt/s 
(see Fig. \ref{fig xrt lc} and \ref{fig swift_xmm_chan_hrd}). 
This threshold  is lower than the equivalent XRT intensity of Obs 1 which was 
believed to be in quiescence. The calculated duty cycle of \ax\
is $\sim51\%$. Due to the random and long separation of the XRT observations, 
the estimated duty cycle may have a large uncertainty.
We used highest 10\% of the X-ray intensities ($\sim$ 0.003 cnt/s) to estimate 
the uncertainty of the threshold. With the threshold in the range of 0.006 $-$ 0.012 
cnt/s,  we estimate a duty cycle in the range of 35\% $-$ 60\%.

The estimated  $\sim$50\% duty cycle indicates that this source is a very active 
DN with comparable quiescence and outburst time. Fig. \ref{fig lc optical 2016}
shows that \ax\ is very active in our new optical observations and also shows
comparable quiescence and outburst time. The new optical light curves 
support the estimated high duty cycle in \ax.

\cite{Britt15} recently measured the duty cycles for an existing sample of 
well-observed, nearby DNe and derived a quantitative empirical relation between 
the duty cycle of DNe outbursts and the X-ray luminosity of the system 
in quiescence. If we assume this relationship also applies to \ax, the estimated
$\sim$50\% duty cycle suggests the X-ray luminosity of this system should be
larger than $10^{32}$ erg/s in quiescence. This suggests that \ax\ might belong 
to a class of DNe with both high average mass transfer rate in outburst and 
high instantaneous accretion rate in quiescence. Assuming the X-ray flux 
of this source  in quiescence is larger than $\sim$1.2$\times10^{-13}$ erg/cm$^2$/s 
(see Table \ref{tab:fit_all}), we could estimate the distance to \ax\ to 
be $\simless$ 1.0 kpc. We note that neither the duty cycle nor the X-ray 
luminosity in quiescence are able to entirely reliably trace the  mass 
accretion rate of the system \citep{Britt15}.

\section{Conclusions}
\label{conclusion}

In this paper we analyse new optical/UV/X-ray light curves and spectra of the 
peculiar CV \ax.  Both the FRED optical outbursts and the lack of apparent 
He II emission lines in the optical spectra  suggest that \ax\ is 
a typical DN. We present  multi-wavelength (optical/UV/X-ray) 
observations of DN \ax\ throughout three mini-outbursts and one normal outburst. 
We find  the UV emission delays the optical emission by $1.0 - 5.4$ days during the rising phase 
of the outburst. The X-ray emission shows suppression during the outburst peak and 
recovery during the end of the outburst.

We also analyse  archival \swift, \chan\ and \xmm\ observations  of 
\ax. We find an approximately anti-correlation between X-ray and UV flux. 
The high-resolution X-ray spectra from \xmm\ and \chan\ can be well 
described either by a single temperature thermal plasma model or by 
an isobaric cooling flow model when its X-ray flux is high. 
We find the maximum temperature of the plasma in quiescence is 
significant higher than that in outburst in \ax.  
Our estimated high duty cycle suggests that the X-ray luminosity of this 
source should be larger than $10^{32}$ erg/s in quiescence.

\section*{Acknowledgements}
This work makes use of optical observations from the Las Cumbres 
Observatory Global Telescope Network, and has made use of the LCOGT 
Archive, which is operated by the California Institute of Technology, 
under contract with the Las Cumbres Observatory.  
The X-ray data are obtained from the High Energy
Astrophysics Science Archive Research Center (HEASARC), provided 
by NASA's Goddard Space Flight Center and NASA's Astrophysics Data 
System Bibliographic Services. We thank Mallory Roberts for useful 
comments and discussions. We thank  B. Sbarufatti, K. L. Page 
and  D. Malesani for approving our \swift\ ToO (target ID 42706) and 
the \swift\ Science  Operations Team for performing the observations.

\appendix
\section{Log of LCOGT 2015 and 2016 observations.}
\label{app-table}
We observed the field of \ax\ with LCOGT suite of 
robotic 1-m telescopes in 2015 and 2016. Some data in 2016 were taken with 
the 2-m Faulkes Telescope South (FTS). All 1-m telescopes are equipped with 
a camera with pixel scale 0.467 arcsec pixel$^{-1}$; this is 0.304 arcsec 
pixel$^{-1}$ for the 2-m data. We adopt a 2.8 arcsec aperture for the 1-m 
data and 1.8 arcsec aperture for the 2-m data, to match the average seeing 
differences between the 1-m and 2-m data. The observation 
logs are provided in Table \ref{tab:optical} and the telescopes are: \\
\\
1m0-03 $=$ Siding Spring, Australia \\
1m0-05 $=$ Cerro Tololo, Chile \\
1m0-10 $=$ SAAO, South Africa \\
1m0-11 $=$ Siding Spring, Australia \\
1m0-12 $=$ SAAO, South Africa \\
1m0-13 $=$ SAAO, South Africa \\
FTS    $=$ 2m telescope at Siding Spring, Australia \\

\begin{table*}
\begin{center}
\caption{Log of LCOGT 2015 and 2016 observations.}
\vspace{-2mm}
\begin{tabular}{llllllll}
\hline
Date UT    & MJD  & magnitude ($R$)  &Telescope  & Filters            &Exposure times (s)\\
\hline
2015-04-30  & 57142.81028 & $17.471 \pm 0.051$   & 1m0-10  & $B,V,R,i^{\prime}$ & 100,100,100,100  \\
2015-05-06  & 57148.06933 & $15.589 \pm 0.014$   & 1m0-05  & $B,V,R,i^{\prime}$ & 200,100,100,100  \\
2015-05-12  & 57154.42698 & $16.530 \pm 0.016$   & 1m0-11  & $B,V,R,i^{\prime}$ & 200,100,100,100  \\
2015-05-25  & 57167.10270 & $17.051 \pm 0.034$   & 1m0-10  & $B,V,R,i^{\prime}$ & 200,100,100,100  \\
2015-06-06  & 57179.70814 & $17.684 \pm 0.062$   & 1m0-12  & $R$		& 100  \\
2015-06-07  & 57180.72358 & $17.285 \pm 0.033$   & 1m0-13  & $R$		& 100  \\
2015-06-10  & 57183.00197 & $15.604 \pm 0.046$   & 1m0-10  & $R$       &  100  \\
2015-06-10  & 57183.00341 & $15.552 \pm 0.059$   & 1m0-10  & $R$       &  100  \\
2015-06-10  & 57183.00483 & $15.535 \pm 0.047$   & 1m0-10  & $R$       &  100  \\
2015-06-10  & 57183.00637 & $15.623 \pm 0.036$   & 1m0-10  & $R$       &  100  \\
2015-06-10  & 57183.00809 & $15.609 \pm 0.038$   & 1m0-10  & $R$       &  100  \\
2015-06-10  & 57183.00970 & $15.592 \pm 0.043$   & 1m0-10  & $R$       & 100  \\
2015-06-10  & 57183.01119 & $15.582 \pm 0.036$   & 1m0-10  & $R$       & 100  \\
2015-06-10  & 57183.01262 & $15.598 \pm 0.051$   & 1m0-10  & $R$       & 100  \\
2015-06-10  & 57183.01405 & $15.528 \pm 0.073$   & 1m0-10  & $R$       & 100  \\
2015-06-10  & 57183.36087 & $15.512 \pm 0.012$   & 1m0-03  & $R$      & 100  \\
2015-06-10  & 57183.36218 & $15.513 \pm 0.012$   & 1m0-03  & $R$      & 100  \\
2015-06-10  & 57183.36349 & $15.499 \pm 0.013$   & 1m0-03  & $R$      & 100  \\
2015-06-10  & 57183.36481 & $15.509 \pm 0.012$   & 1m0-03  & $R$      & 100  \\
2015-06-10  & 57183.36612 & $15.509 \pm 0.012$   & 1m0-03  & $R$      & 100  \\
2015-06-10  & 57183.36744 & $15.502 \pm 0.013$   & 1m0-03  & $R$      & 100  \\
2015-06-10  & 57183.36875 & $15.514 \pm 0.011$   & 1m0-03  & $R$      & 100  \\
2015-06-10  & 57183.37007 & $15.489 \pm 0.011$   & 1m0-03  & $R$      & 100  \\
2015-06-10  & 57183.37138 & $15.501 \pm 0.010$   & 1m0-03  & $R$      & 100  \\
2015-06-10  & 57183.37270 & $15.509 \pm 0.010$   & 1m0-03  & $R$      & 100  \\
2015-06-10  & 57183.99798 & $15.527 \pm 0.010$   & 1m0-05  & $R$      	& 100  \\
2016-02-01  & 57419.66520 & $16.796 \pm 0.040$   & 1m0-03 & $R$ & 200 \\
2016-02-01  &  57419.69488 & $16.764 \pm 0.017$   & 1m0-03 & $R$ & 200 \\
2016-02-03  &  57421.08700 & $17.022 \pm 0.092$   & 1m0-10 & $R$ & 200 \\
2016-02-04  &  57422.29909 & $16.981 \pm 0.017$   & 1m0-05 & $R$ & 200 \\
2016-02-05  &  57423.33328 & $17.041 \pm 0.025$   & 1m0-05 & $R$ & 200 \\
2016-02-06  &  57424.03382 & $17.124 \pm 0.022$   & 1m0-13 & $R$ & 200 \\
2016-02-07  &  57425.03107 & $17.118 \pm 0.022$   & 1m0-13 & $R$ & 200 \\
2016-02-07  &  57425.72135 & $17.146 \pm 0.024$   & FTS & $R$ & 120 \\
2016-02-08  &  57426.02834 & $16.963 \pm 0.019$   & 1m0-10 & $R$ & 200 \\
2016-02-09  &  57427.02551 & $17.095 \pm 0.038$   & 1m0-13 & $R$ & 200 \\
2016-02-09  &  57427.68389 & $17.026 \pm 0.033$   & FTS & $R$ & 120 \\
2016-02-10  &  57428.02394 & $17.026 \pm 0.021$   & 1m0-13 & $R$ & 200 \\
2016-02-10  &  57428.63314 & $16.994 \pm 0.020$   & FTS & $R$ & 120 \\
2016-02-11  &  57429.04155 & $16.992 \pm 0.019$   & 1m0-13 & $R$ & 200 \\
2016-02-12  &  57430.02881 & $16.947 \pm 0.019$   & 1m0-10 & $R$ & 200 \\
2016-02-12  &  57430.62836 & $16.880 \pm 0.019$   & FTS & $R$ & 120 \\
2016-02-13  &  57431.11716 & $16.795 \pm 0.034$   & 1m0-13 & $R$ & 200 \\
2016-02-14  &  57432.01188 & $16.757 \pm 0.017$   & 1m0-13 & $R$ & 200 \\
2016-02-14  &  57432.69475 & $16.691 \pm 0.016$   & FTS & $R$ & 120 \\
2016-02-15  &  57433.26795 & $16.607 \pm 0.013$   & 1m0-05 & $R$ & 200 \\
2016-02-15  &  57433.62776 & $16.642 \pm 0.023$   & FTS & $R$ & 120 \\
2016-02-16  &  57434.26027 & $16.590 \pm 0.013$   & 1m0-05 & $R$ & 200 \\
2016-02-16  &  57434.68099 & $16.562 \pm 0.022$   & FTS & $R$ & 120 \\
2016-02-17  &  57435.27446 & $16.713 \pm 0.014$   & 1m0-05 & $R$ & 200 \\
2016-02-17  &  57435.67131 & $16.947 \pm 0.079$   & FTS & $R$ & 120 \\
2016-02-18  &  57436.01093 & $16.803 \pm 0.016$   & 1m0-10 & $R$ & 200 \\
2016-02-18  &  57436.34304 & $16.918 \pm 0.015$   & 1m0-05 & $R$ & 200 \\
2016-02-18  &  57436.61226 & $17.040 \pm 0.025$   & FTS & $R$ & 120 \\
2016-02-19  &  57437.25208 & $17.138 \pm 0.024$   & 1m0-05 & $R$ & 200 \\
2016-02-19  &  57437.65759 & $17.195 \pm 0.026$   & FTS & $R$ & 120 \\
2016-02-20  &  57438.25358 & $17.264 \pm 0.030$   & 1m0-05 & $R$ & 200 \\
2016-02-21  &  57439.29170 & $17.274 \pm 0.032$   & 1m0-05 & $R$ & 200 \\
2016-02-22  &  57440.00040 & $17.559 \pm 0.069$   & 1m0-13 & $R$ & 200 \\
2016-02-23  &  57441.01024 & $17.417 \pm 0.046$   & 1m0-13 & $R$ & 200 \\
2016-02-24  &  57442.00571 & $17.492 \pm 0.047$   & 1m0-13 & $R$ & 200 \\
2016-02-25  &  57443.02726 & $17.485 \pm 0.044$   & 1m0-13 & $R$ & 200 \\
\hline
\end{tabular}
\normalsize
\label{tab:optical}
\end{center}
\end{table*}
\addtocounter{table}{-1}
\begin{table*}{-1}
\begin{center}
\caption{Continued.}
\vspace{-2mm}
\begin{tabular}{llllllll}
\hline
Date UT     &   MJD        & magnitude ($R$)      &Telescope  & Filters            &Exposure times (s)\\
\hline
2016-02-27  &  57445.62117 & $17.318 \pm 0.053$   & 1m0-03 & $R$ & 200 \\
2016-02-28  &  57446.01335 & $17.420 \pm 0.035$   & 1m0-13 & $R$ & 200 \\
2016-03-01  &  57448.06750 & $17.008 \pm 0.018$   & 1m0-10 & $R$ & 200 \\
2016-03-01  &  57448.62489 & $17.028 \pm 0.021$   & 1m0-03 & $R$ & 200 \\
2016-03-03  &  57450.60801 & $17.112 \pm 0.018$   & 1m0-03 & $R$ & 200 \\
2016-03-04  &  57451.60506 & $17.130 \pm 0.018$   & 1m0-03 & $R$ & 200 \\
2016-03-05  &  57452.60201 & $17.313 \pm 0.021$   & 1m0-03 & $R$ & 200 \\
2016-03-07  &  57454.40306 & $17.337 \pm 0.024$   & 1m0-05 & $R$ & 200 \\
2016-03-07  &  57454.62370 & $17.337 \pm 0.022$   & 1m0-03 & $R$ & 200 \\
2016-03-09  &  57456.36274 & $17.483 \pm 0.023$   & 1m0-05 & $R$ & 200 \\
2016-03-09  &  57456.94641 & $17.457 \pm 0.024$   & 1m0-13 & $R$ & 200 \\
2016-03-10  &  57457.58832 & $17.391 \pm 0.025$   & 1m0-03 & $R$ & 200 \\
2016-03-11  &  57458.58716 & $17.275 \pm 0.020$   & 1m0-03 & $R$ & 200 \\
2016-03-12  &  57459.63460 & $17.106 \pm 0.019$   & 1m0-03 & $R$ & 200 \\
2016-03-13  &  57460.74649 & $16.931 \pm 0.020$   & 1m0-03 & $R$ & 200 \\
2016-03-14  &  57461.63193 & $16.717 \pm 0.014$   & 1m0-03 & $R$ & 200 \\
2016-03-15  &  57462.96572 & $16.552 \pm 0.025$   & 1m0-13 & $R$ & 200 \\
2016-03-16  &  57463.93002 & $16.539 \pm 0.015$   & 1m0-13 & $R$ & 200 \\
2016-03-17  &  57464.56933 & $16.648 \pm 0.017$   & 1m0-03 & $R$ & 200 \\
2016-03-18  &  57465.61511 & $16.686 \pm 0.026$   & 1m0-03 & $R$ & 200 \\
2016-03-20  &  57467.91649 & $17.066 \pm 0.033$   & 1m0-13 & $R$ & 200 \\
2016-03-21  &  57468.00108 & $17.096 \pm 0.025$   & 1m0-10 & $R$ & 200 \\
2016-03-22  &  57469.16480 & $17.188 \pm 0.039$   & 1m0-05 & $R$ & 200 \\
2016-03-23  &  57470.15458 & $17.210 \pm 0.031$   & 1m0-13 & $R$ & 200 \\
2016-03-24  &  57471.14439 & $17.373 \pm 0.049$   & 1m0-13 & $R$ & 200 \\
2016-03-26  &  57473.64500 & $17.209 \pm 0.035$   & 1m0-03 & $R$ & 200 \\
2016-03-27  &  57474.74986 & $17.387 \pm 0.035$   & 1m0-03 & $R$ & 200 \\
2016-03-28  &  57475.26752 & $17.307 \pm 0.033$   & 1m0-05 & $R$ & 200 \\
2016-03-29  &  57476.02154 & $17.393 \pm 0.057$   & 1m0-13 & $R$ & 200 \\
2016-03-29  &  57476.37488 & $17.288 \pm 0.036$   & 1m0-05 & $R$ & 200 \\
2016-03-30  &  57477.36813 & $17.600 \pm 0.040$   & 1m0-05 & $R$ & 200 \\
2016-04-02  &  57480.52608 & $17.793 \pm 0.033$   & 1m0-03 & $R$ & 200 \\
2016-04-03  &  57481.13232 & $17.424 \pm 0.026$   & 1m0-05 & $R$ & 200 \\
2016-04-03  &  57481.13548 & $17.550 \pm 0.030$   & 1m0-05 & $R$ & 200 \\
2016-04-04  &  57482.04207 & $17.424 \pm 0.045$   & 1m0-13 & $R$ & 200 \\
2016-04-05  &  57483.51738 & $15.862 \pm 0.009$   & 1m0-03 & $R$ & 200 \\
2016-04-07  &  57485.12103 & $15.458 \pm 0.007$   & 1m0-05 & $R$ & 200 \\
2016-04-07  &  57485.12456 & $15.460 \pm 0.007$   & 1m0-05 & $R$ & 200 \\
2016-04-07  &  57485.14465 & $15.440 \pm 0.007$   & 1m0-05 & $R$ & 200 \\
2016-04-07  &  57485.16703 & $15.451 \pm 0.007$   & 1m0-05 & $R$ & 200 \\
2016-04-07  &  57485.17373 & $15.451 \pm 0.008$   & 1m0-05 & $R$ & 200 \\
2016-04-07  &  57485.34353 & $15.399 \pm 0.007$   & 1m0-05 & $R$ & 200 \\
2016-04-07  &  57485.41653 & $15.415 \pm 0.007$   & 1m0-05 & $R$ & 200 \\
2016-04-07  &  57485.51197 & $15.424 \pm 0.011$   & 1m0-03 & $R$ & 200 \\
2016-04-07  &  57485.51545 & $15.426 \pm 0.013$   & 1m0-03 & $R$ & 200 \\
2016-04-08  &  57486.13174 & $15.170 \pm 0.007$   & 1m0-05 & $R$ & 200 \\
2016-04-08  &  57486.19285 & $15.132 \pm 0.006$   & 1m0-05 & $R$ & 200 \\
2016-04-08  &  57486.86447 & $15.173 \pm 0.007$   & 1m0-13 & $R$ & 200 \\
2016-04-09  &  57487.17049 & $15.152 \pm 0.006$   & 1m0-05 & $R$ & 200 \\
2016-04-11  &  57489.93411 & $15.247 \pm 0.007$   & 1m0-10 & $R$ & 200 \\
2016-04-12  &  57490.49836 & $15.307 \pm 0.008$   & 1m0-03 & $R$ & 200 \\
2016-04-13  &  57491.17433 & $15.313 \pm 0.006$   & 1m0-05 & $R$ & 200 \\
2016-04-15  &  57493.49691 & $15.665 \pm 0.012$   & 1m0-03 & $R$ & 200 \\
2016-04-16  &  57494.16704 & $15.848 \pm 0.010$   & 1m0-13 & $R$ & 200 \\
2016-04-18  &  57496.48241 & $16.362 \pm 0.017$   & 1m0-03 & $R$ & 200 \\
2016-04-19  &  57497.30683 & $16.587 \pm 0.018$   & 1m0-05 & $R$ & 200 \\
2016-04-24  &  57502.85855 & $17.373 \pm 0.035$   & 1m0-10 & $R$ & 200 \\
2016-04-24  &  57502.88819 & $17.508 \pm 0.046$   & 1m0-13 & $R$ & 200 \\
2016-04-26  &  57504.85539 & $18.029 \pm 0.052$   & 1m0-10 & $R$ & 200 \\
2016-04-27  &  57505.50374 & $17.827 \pm 0.065$   & 1m0-03 & $R$ & 200 \\
2016-04-27  &  57505.87528 & $17.951 \pm 0.046$   & 1m0-13 & $R$ & 200 \\
2016-04-30  &  57508.80454 & $17.788 \pm 0.037$   & 1m0-13 & $R$ & 200 \\
\hline
\end{tabular}
\normalsize
\label{tab:optical-2}
\end{center}
\end{table*}

\end{document}